# A Study on Feature Selection Techniques in Educational Data Mining

M. Ramaswami and R. Bhaskaran

**Abstract**—Educational data mining (EDM) is a new growing research area and the essence of data mining concepts are used in the educational field for the purpose of extracting useful information on the behaviors of students in the learning process. In this EDM, feature selection is to be made for the generation of subset of candidate variables. As the feature selection influences the predictive accuracy of any performance model, it is essential to study elaborately the effectiveness of student performance model in connection with feature selection techniques. In this connection, the present study is devoted not only to investigate the most relevant subset features with minimum cardinality for achieving high predictive performance by adopting various filtered feature selection techniques in data mining but also to evaluate the goodness of subsets with different cardinalities and the quality of six filtered feature selection algorithms in terms of F-measure value and Receiver Operating Characteristics (ROC) value, generated by the NaïveBayes algorithm as base-line classifier method. The comparative study carried out by us on six filter feature section algorithms reveals the best method, as well as optimal dimensionality of the feature subset. Benchmarking of filter feature selection method is subsequently carried out by deploying different classifier models. The result of the present study effectively supports the well known fact of increase in the predictive accuracy with the existence of minimum number of features. The expected outcomes show a reduction in computational time and constructional cost in both training and classification phases of the student performance model.

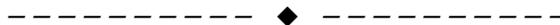

## 1 INTRODUCTION

Educational Data Mining, which is an area of scientific inquiry centered on the development of methods not only for making discoveries within the unique kinds of data that come from educational settings but also for using those methods subsequently to understand effectively the students and the settings which they learn in, has emerged as an independent research area in recent years [1]. One of the key areas of applications of EDM is improvement of student models that would predict student's characteristics or academic performances in schools, colleges and other educational institutions. Prediction of student performance with high accuracy is useful in many contexts in all educational institutions for identifying slow learners and distinguishing students with low academic achievement or weak students who are likely to have low academic achievements. The end product of models would be beneficial to the teachers, parents and educational planners not only for informing the students during their study, whether their current behavior could be associated with positive and negative outcomes of the past, but also for providing advice to rectify problems. As the end products of the models would be presented regularly to students in a comprehensive form, these end products would facilitate reflection and self-regulation during their study.

In prediction, the goal is to develop a model which can infer a single aspect of the data (dependent variable) from some combination of other aspects of the data (independent variables). In fact, prediction requires having labels for the output variable for a limited data set, where a label represents some trusted "ground truth" information about the output variable's value in specific cases. At the same time, prediction has two key uses within educational data mining. As far as first type of usage is concerned, prediction methods can be used to study the features of a model that are important for prediction and giving information about the underlying construction. This is a common approach in programs of research that attempt to predict student educational outcomes, without predicting intermediate or mediating factors first. In a second type of usage, prediction methods are used in order to predict the output value that would be in context and obtain a label for that construction.

The performance of prediction model highly depends on the choice of selection of most relevant variable from the list of variables used in student data set. This can be achieved by means of applying different feature selection techniques on data set. In fact, percentage of accuracy is generally not preferred for classification, as values of accuracy are highly dependent on the base rates of different classes. It is worth mentioning here that Receiver-Operating Characteristics (ROC) value and F-Measure can be used for assessing the goodness of a predictor.

The academic achievement of higher secondary school education in India is a turning point in the life of any student, as it serves as a very important link between the higher and higher secondary education of students. But,

————————————————
- *M. Ramaswami is with Department of Computer Science, Madurai Kamaraj University, Madurai, India, 625 021.*
- *R. Bhaskaran is with School of Mathematics, Madurai Kamaraj University, Madurai, Tamil Nadu,, India, 625 021.*

there are determinants like demographic, academic and socio-economic factors of students that restrict the students' performance. This necessitates the need for some forecasting system to predict the academic performance of students at plus two examinations. This is an attempt made for the first time in this aspect, which is mainly devoted to design and develop a prediction model by taking into account variables pertaining to the Indian society, for Indian educational system. In this connection, the objectives are (i) investigation on the most relevant subset features with minimum cardinality for achieving high predictive performance by adopting various filtered feature selection techniques in data mining and (ii) evaluation of goodness of subsets with different cardinalities and the quality of six filtered feature selection algorithms in terms of F-measure value and Receiver Operating Characteristics (ROC) value, generated by the Naïve-Bayes algorithm as base-line classifier method.

## 2 FEATURE SELECTION

Feature selection has been an active and fruitful field of research area in pattern recognition, machine learning, statistics and data mining communities [2, 3]. The main objective of feature selection is to choose a subset of input variables by eliminating features, which are irrelevant or of no predictive information. Feature selection has proven in both theory and practice to be effective in enhancing learning efficiency, increasing predictive accuracy and reducing complexity of learned results [4, 5]. Feature selection in supervised learning has a main goal of finding a feature subset that produces higher classification accuracy.

As the dimensionality of a domain expands, the number of features N increases. Finding an optimal feature subset is intractable and problems related feature selections have been proved to be NP-hard [6]. At this juncture, it is essential to describe traditional feature selection process, which consists of four basic steps, namely, subset generation, subset evaluation, stopping criterion, and validation [7]. Subset generation is a search process that produces candidate feature subsets for evaluation based on a certain search strategy. Each candidate subset is evaluated and compared with the previous best one according to a certain evaluation. If the new subset turns to be better, it replaces best one. This process is repeated until a given stopping condition is satisfied.

Ranking of features determines the importance of any individual feature, neglecting their possible interactions. Ranking methods are based on statistics, information theory, or on some functions of classifier's outputs [8]. Algorithms for feature selection fall into two broad categories namely wrappers that use the learning algorithm itself to evaluate the usefulness of features and filters that evaluate features according to heuristics based on general characteristics of the data [7].

Several justifications for the use of filters for subset selection have been discussed [6, 9, 10] and it has been reported that filters are comparatively faster than wrappers. Many student performance prediction models have been proposed and comparative analyses of different classifier models using Decision Tree, Bayesian Network, and other classification algorithms have also been discussed ([11], [12], [13], [14]). But, they reveal only classifier accuracy without performing the feature selection procedures.

## 3 DATA SOURCE AND PREDICTION OUTCOMES

School education in India is a two-tier system, the first ten years covering general education followed by two years of senior secondary education. This two-year education, which is also known as Higher Secondary Education, is important because it is a deciding factor for opting desired subjects of study in Higher education. In fact, the higher secondary education acts as a bridge between school education and the higher learning specializations that are offered by colleges and universities. At this juncture, it is essential to measure the academic performance of students, which is a challenging task due to the contributions of socio-economic, psychological and environmental factors. Measurement of academic performance is carried out by using the predictive models and it is to be noted that prediction modeling process are composed of a feature extraction, which aims at preserving most of the relevant features of student's characteristics while evading any source of adverse variability, and a classification stage that identifies the feature vector with appropriate class. Incidentally, the classification operation based on the probability density function [15] of the feature vector space is ineffective in the case of inappropriate choice of features, or in the presence of parameters, which do not give useful information. Thus, combining the classification process with feature selection technique has become a necessity in the model construction [16].

The main source of data for this study is the responses obtained from students through a questionnaire with close-end questions. The responces give demographic details, family details, socio-economic details, previous academic performance at secondary level from different schools and other environmental details. A total of 1969 higher secondary students from different schools in different districts of the state Tamil Nadu, India, supplied the details. We notice that the original feature vector of student performance data consisted of 32 features that were predictive variables. Besides, there was a two-case class variable *result* (*pass / fail*), which was considered as response variable. All these predictive and responsive variables shown in Table 1 belonged to the type of nominal data.

Feature selection is normally done by searching the space of attribute subsets, evaluating each one. This is achieved by combining attribute subset evaluator with a search method. In the present investigation, an evaluation of six filter feature subset methods with rank search or Greedy search method was performed to find out the best feature sets and they are listed below for reference:

TABLE 1 LIST OF FEATURES USED IN STUDY

| Sl.No. and Variable Name | Description |
|---|---|
| (1) SEX | student's sex |
| (2) ESight | student's eye vision |
| (3) Comm | student's community |
| (4) PHD | physically handicapped |
| (5) FHBT | student's food habit |
| (6) FAM-Size | student's family size |
| (7) LArea | student's living area |
| (8) No-EB | number of elder brothers |
| (9) No-ES | number of elder sisters |
| (10) No-YB | number of younger brothers |
| (11) No-YS | number of younger sisters |
| (12) JIFamily | student's family status |
| (13) TransSchool | mode of transportation to school |
| (14) Veh-Home | own vehicle at home |
| (15) PSEdu | student had primary education |
| (16) ESEdu | type of institution at elementary level |
| (17) StMe | type of secondary syllabus |
| (18) XMark-Grade | marks/grade obtained at secondary level |
| (19) TYP-SCH | type of school |
| (20) LOC-SCH | location of school |
| (21) MED | medium of Instruction |
| (22) PTution | private tuition- No. Subjects |
| (23) SPerson | sports/athletic |
| (24) SpIndoor | type of indoor game |
| (25) SpOutdoor | type of outdoor game |
| (26) Cstudy | care of study at home |
| (27) FEDU | father's education |
| (28) FOCC | father's occupation |
| (29) FSAL | father's monthly income |
| (30) MEDU | mother's education |
| (31) MOCC | mother's occupation |
| (32) MSAL | mother's monthly income |
| (Response Variable) | |
| (33) HSCGrade | marks/grade obtained at HSc Level |

1) Correlation-based Attribute evaluation (**CB**),
2) Chi-Square Attribute evaluation (**CH**),
3) Gain-Ratio Attribute evaluation (**GR**),
4) Information-Gain Attribute evaluation(**IG**),
5) Relief Attribute evaluation (**RF**) and
6) Symmetrical Uncertainty Attribute evaluation (**SU**)

These entire filter techniques mentioned above could assess the relevance of features [16] on the basis of the intrinsic properties of the data. Feature selection often increases classifier efficency through the reduction of the size of the effective features. Therefore, there is a need to verify the relevance of all the features in the feature vector. In this connection, we performed all the above six feature selection techniques based on different measures to choose the best subsets for a given cardinality. We used subsequently the NaiveBayes classification algorithm, which is one of the simplest instances of probabilistic classifiers with accurate outcomes as that of state-of-the-art learning algorithms for prediction model construction, as a base line classifier to select the final best subset among the best subsets across different cardinalities. The measures like ROC Values and Macro-Average F1-Measure values are used in the present investigation.

## 4 RESULTS AND DISCUSSION

The present investigation focuses on various feature selection techniques, which is one of the most important and frequently used in data preprocessing for data mining. The general procedures on Feature Selection in terms of Filter method is followed with the effect of feature selection techniques on a generated database on higher secondary students. Effectiveness of the algorithms is presented in terms of different measures like ROC Values and F1-Measure values.

Initially all feature selection methods were applied on the original feature set and the features were ranked according to their merits in ascending order. Since no agreement was found among the feature ranking methods, we performed student performance evaluation in terms of ROC value and F1-Measure values on multiple subsets of feature vectors. In fact, the evaluation on the basis of ROC value and F1-Measure was carried out iteratively on the multiple subsets starting from two with one as an increment from the ranking list.

The Receiver Operating Characteristics (ROC) curve is a graphical representation of the trade off between the false negative and false positive rates for every possible cut off. Equivalently, the ROC value is the representation of the tradeoffs between Sensitivity and Specificity. The evaluation measures with variations of ROC values and F1-Mesures are generated from an Open Source Data Mining suite, WEKA [17] that offers a comprehensive set of state-of-the-art machine learning algorithms as well as set of autonomous feature selection and ranking methods. The generated evaluation measures are shown in Fig. 1 for reference.

While the X-axis represents the number of features,

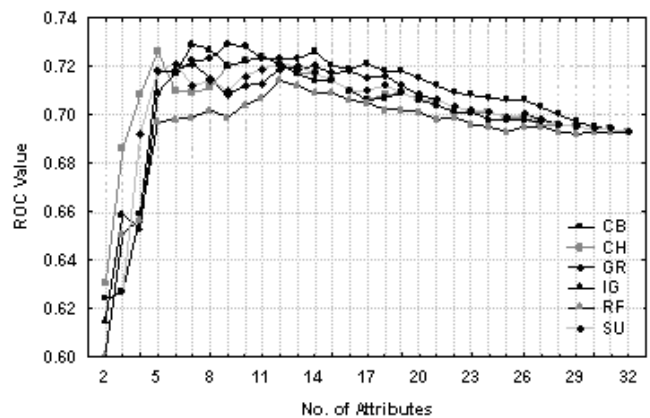

Fig.1. Number of Attributes Vs ROC value

the Y-axis represents the ROC value for each feature subset generated by six filter features. The maximum ROC value of all algorithms and the corresponding cardinalities from the Fig.1 are presented in Table 2. This is quite useful for fixing the optimal size of the feature subsets with the highest ROC values.

We observe from Table 2 that the ROC values are the highest for CB and IG techniques. Although both these techniques namely CB and IG resulted in the ROC value of 0.729, we note that **IG** could attain the maximum ROC value when it had 7 features. But, it was also observed that **CB** could attain the maximum ROC value when it had 9 features. So, we deduce that IG has the optimal dimensionality in the dataset of higher secondary students.

TABLE 2
PEAK ROC VALUES

| FST | CB | CH | GR | IG | RF | SU |
|---|---|---|---|---|---|---|
| No. of Attributes | **9** | 5 | 7 | **7** | 12 | 6 |
| ROC Value | **0.729** | 0.726 | 0.721 | **0.729** | 0.714 | 0.721 |

TABLE 3
PEAK VALUES OF F1-MEASURE

| FST | CB | CH | GR | IG | RF | SU |
|---|---|---|---|---|---|---|
| No. of Attributes | 17 | **12** | 12 | **12** | 12 | **12** |
| F1-Measure | 0.584 | **0.592** | 0.588 | **0.592** | 0.582 | **0.592** |

F1- Measure, which is another measure for evaluating the effectiveness of feature selection techniques, is the harmonic mean of the precision and recall and this metric is more relevant for evaluating the performance of the classifier. The generated macro-averaged F-measures, which are presented in Fig. 2, could show each of the fea-ture selection metrics with variations in the number of features.

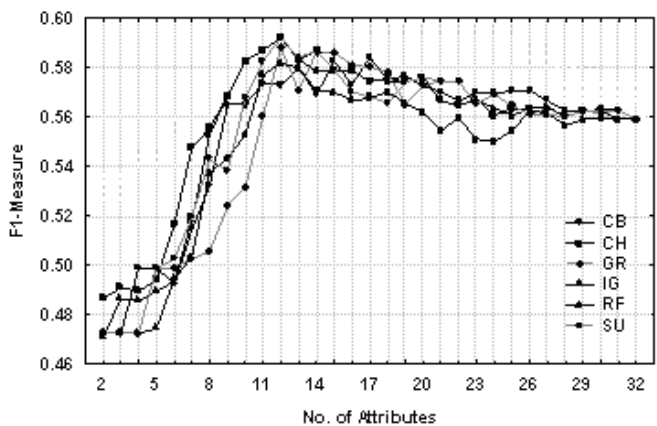

Fig. 2. Number of attributes Vs F1-Measure values

While the X-axis represents the number of features, the Y-axis represent the F1- Measure for each feature subset generated by six filter feature techniques. It is worth mentioning here that the overall trends and the separation of the top performing curves are of interest but not of the absolute values. The maximum F1-Measure of all algorithms and the corresponding cardinalities from the Fig. 2 are presented in Table 3. This is quit useful for fixing the optimal size of the feature subsets with the highest F1-Measures.

We observe from Table 3 that the F1-Measures are the highest for CH, IG and SU techniques. All these techniques namely CH, IG and SU resulted in the F1-Measure of 0.592 with 12 features for their best performance. So, we deduce that CH, IG and SU have the optimal dimensionality in the dataset of higher secondary students.

The ROC values are the highest for CB and IG techniques with 9 and 7 features respectively, whereas the F1-measures are the highest for CH, IG and SU techniques with 12 features each. Benchmarking is essential now for arriving at the best feature selection algorithm and conclusions can be drawn on the basis of classification algorithms. Predictive models constructed in terms of four supervised learnining algorithms, *viz*. NaiveBayes, Voted Perceptron, OneR and PART enable us to obtain the efficency of the five optimal subsets *viz*. CB-9, IG-7, CH-12, IG-12, and SU-12. The top ranking features that are produced by CB-9, IG-7, CH-12, IG-12, and SU-12 algorithms for further predictive analysis are presented in Table 4.

A NaiveBayes classifier can achieve relatively good performance on classification tasks [19], based on the elementary Bayes' Theorem. It greatly simplifies learning by assuming that features are independent for a given class variable. The voted-perceptron algorithm is a margin maximizing algorithm based on an iterative application of the classic perceptron algorithm. Using voted perceptrons one can get comparable performance with SVMs [18] but learning is much faster. A OneR is a simple, cheap method that often comes up with quite good rules for characterizing the structure in data. It can generate a one-level decision tree expressed in the form of a set of rules that test one particular attribute. PART obtains rules from the partial decision trees and it builds the tree using C4.5's heuristics. All these clas-sifiers are used for the present investigation so as to fix the optimal subset.

TABLE 4
TOP RANKING FEATURE SUBSETS IN CB, IG,

| Top ranking attribute numbers based on ROC values | CB-9 | 17,1,21,18,20,32,13,7,28 |
|---|---|---|
| | IG-7 | 18,17,28,21,20,13,1 |
| Top ranking attribute numbers based on F1-Measure | CH-12 | 18,28,20,1,21,19,7,13,17,27,32,22 |
| | IG-12 | 18,17,28,21,20,13,1,7,19,27,32,22 |
| | SU-12 | 17,21,1,20,18,7,19,13,22,28,32,27 |

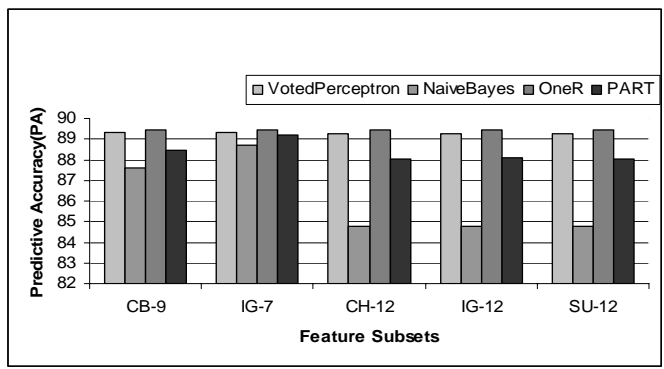

Fig. 3. Performance of 4 classifiers over a different number/subset of features

The results of the comparative study of four different classifiers carried out against feature subsets generated by the five different feature selection procedures are shown in Fig. 3. Clearly all the four classifiers performed well for the feature subset (IG-7) generated by information gain values. Also the classifiers Voted Perceptron, OneR showed high predictive accuracy which is more than 89 percent. In particular, the Voted Perceptron steadily provided the same level of predictive performance for all the feature subsets.

We observe from the present insvestigation that, a judicious combination of ROC values, F1-Measure, high Predictive Accuracy (PA) and low Root Mean Square Error (RMSE) for the IG method with top 7 features (i.e., **IG-7**), yields an optimal dimensionality of the feature set.

## 5 CONCLUSION

In this paper, we carried out a comparative study of six filter feature section algorithms by which we could reach the best method as well as optimal dimensionality of the feature subset. Benchmarking of filter feature selection method was subsequently carried out by deploying different classifier models. The results of the present study effectively supported the well known fact of increase in the predictive accuracy with the existence of minimum number of features. The expected outcomes show a reduction in computational time and constructional cost in both training and classification phases of the student performance model.

**M. Ramaswami** is Associate Professor in the Department of Computer Science, Madurai Kamaraj University, Tamil Nadu, India. He obtained his M.Sc. (Maths) from Bharathiyar University, M.C.A from Madurai Kamaraj University and M.Phil. in Computer Applications from Manonmaniam Sundaranar University. He is currently doing research in Data Mining and Knowledge Discovery.

**R. Bhaskaran** received his M.Sc. (Mathematics) from IIT, Chennai, India (1973) and obtained his Ph.D. in Mathematics from the University of Madras (1980). He then joined the School of Mathematics, Madurai Kamaraj University as Lecturer. At present, he is working as Professor of Mathematics. His research interests include non-archimedean analysis, Image Processing, Data Mining and Software development for learning Mathematics.